\documentclass[12pt]{article}
\usepackage{graphicx}
\usepackage{amsmath}
\usepackage{amssymb}
\def\ltap{\raisebox{-.4ex}{\rlap{$\sim$}} \raisebox{.4ex}{$<$}}
\def\gtap{\raisebox{-.4ex}{\rlap{$\sim$}} \raisebox{.4ex}{$>$}}

\def\sinthwsp {${\rm sin}^2 \theta_W\ $}

\def\sinthlsp {${\rm sin}^2 \theta_W^{\ell\ {\rm eff}}\ $}

\def\sth34{$|{\rm sin}\theta_{34}|$}
\def\sth34sp{$|{\rm sin}\theta_{34}|\ $}

\def\alr {$A_{LR}$}
\def\alrsp {$A_{LR}\ $}
\def\afbb {$A_{FB}^b$}
\def\afbbsp {$A_{FB}^b\ $}

\def\afbc {$A_{FB}^c$}

\def\qfb {$Q_{FB}$}

\def\mh {$m_H$}
\def\mhsp {$m_H$\ }
\def\chisqsp {$\chi^2$\ }
\def\chisq {$\chi^2$}

\def\th34{$\theta_{34}$}
\def\th34sp{$\theta_{34}\ $}
\def\almz{$\alpha(m_Z)$}

\def\thcabsp{$\theta_{\rm Cabbibo}\ $}

\def\dalfive{$\Delta \alpha^{(5)}$}
\def\dalfivesp{$\Delta \alpha^{(5)}\ $}
\def\als{$\alpha_S$}%
\def\alssp{$\alpha_S\ $}
\def\journal{\topmargin 0.0in   \oddsidemargin 0in
        \headheight 0pt \headsep 0pt
        \textwidth 6.5in 
\textheight 9in 
        \marginparwidth 1.5in
        \parindent 2em
        \parskip .5ex plus .1ex         \jot = 1.5ex}
\journal

\begin{document}
\begin{titlepage}

\noindent June 25, 2010 \\
\noindent Revised July 27, 2010

\begin{center}

\vskip .5in

{\large Higgs Mass Constraints on a Fourth Family:\\
Upper and Lower Limits on CKM Mixing}

\vskip .5in

Michael S. Chanowitz

\vskip .2in

{\em Theoretical Physics Group\\
     Lawrence Berkeley National Laboratory\\
     University of California\\
     Berkeley, California 94720}
\end{center}

\vskip .25in

\begin{abstract}

Limits on the Higgs boson mass restrict CKM mixing of a possible
fourth family beyond the constraints previously obtained from
precision electroweak data alone. Existing experimental and
theoretical bounds on \mhsp already significantly restrict the allowed
parameter space.  Zero CKM mixing is excluded and mixing of order
\thcabsp is allowed.  Upper and lower limits on 3-4 CKM mixing are
exhibited as a function of \mh. We use the default inputs of the
Electroweak Working Group and also explore the sensitivity of both the
three and four family fits to alternative inputs.

\end{abstract}

\end{titlepage}

\newpage

\renewcommand{\thepage}{\arabic{page}}
\setcounter{page}{1}

{\it \noindent \underline{Introduction}} 

A fourth family of quarks and leptons will be easy to discover or
exclude at the LHC\cite{fhs,lhcsearch} if the quark masses lie within
the $m_Q \ \ltap \ 500$ GeV domain defined by perturbative partial
wave unitarity,\cite{cfh} and even if they are too heavy to observe
directly they will induce a large signal in $gg \to ZZ$ that will be
clearly visible at the LHC.\cite{mczz} A fourth family could be the
key to many unsolved puzzles, such as the hierarchies of the fermion
mass spectrum\cite{flavor} including neutrino masses and
mixing,\cite{nu} electroweak symmetry breaking,\cite{ewsb}
baryogenesis,\cite{ewbaryogenesis} and a variety of interesting
phenomena in CP and flavor physics.\cite{flavor-cp} The four family
Standard Model is likely to have a low cutoff above which new dynamics
would emerge, which could be as low as $\sim 1{1\over 2}$ to $2\ {\rm
  TeV}$ if the quark masses are near the perturbative unitarity
limit.\cite{holdomBSM4,hashimoto}

It has been known for a while that SM4, the four family Standard
Model, is consistent with the precision electroweak (PEW)
data,\cite{he,okun,kribs} as confirmed recently by two independent
global fits.\cite{mc1,el} While SM4 does not greatly improve the
quality of the fit (our best SM4 fit has \chisqsp 1.4 units lower than
SM3), it can, as first noted in \cite{okun}, resolve the tension with
the LEPII 114 GeV lower limit on the Higgs boson mass that is
especially acute if the \afbbsp anomaly is attributed to
underestimated systematic error.\cite{mcafbb} Primarily because of the
\afbbsp anomaly, the Standard Model fit presented below, using EWWG
inputs\cite{ewwgphysrpt} (except $\Gamma_W$ as noted below), has just
a 14\% confidence level, which can only be appreciably improved by new
physics models with flavor nonuniversal interactions. As a result few
of the new physics models under active consideration are able to raise
the confidence level significantly.

In previous work we showed that CKM mixing of the fourth family is
most effectively constrained by nondecoupling contributions to the
$\rho$ (or $T$) parameter, proportional to fourth family mixing angles
and masses.\cite{mc1} To constrain SM4 or any other BSM scenario, it
is not sufficient to consider BSM perturbations around the SM best fit,
but rather it is essential to perform global fits that vary both the
SM and BSM parameters, since the best fit may occur at values of the
SM parameters (especially \mh) that are quite different from their
values in the SM best fit.  In the previous work we followed the
default procedures of the EWWG\cite{ewwgphysrpt}, including their data
set, input parameters, and experimental correlations, implemented via
the ZFITTER code\cite{zfitter} with two loop EW radiative
corrections.\cite{twoloop} As expected our SM3 fit agrees very closely
with the EWWG fit\cite{ewwgs09} (see below). For SM4 we found that
fourth family CKM mixing of order \thcabsp is allowed, leaving room
for an SM4 explanation of possible flavor puzzles in the existing
data. Our results have recently been confirmed by a second study using
the same methodology.\cite{houzfitter}

In this paper we incorporate recently obtained constraints on the SM4
Higgs boson mass into the PEW analysis. Because of the large
enhancement of $gg \to H \to WW$ in SM4, CDF and D0 have been able to
exclude the SM4 Higgs boson at 95\% CL for $131 \leq m_H \leq 204$
GeV.\cite{tevatronlimit} As shown below this constraint combined with
the EW fit and the LEPII limit on \mhsp excludes a large portion of
the SM4 parameter space with small or vanishing fourth family CKM
mixing.  A potentially stronger, theoretical constraint follows from
the RG/stability analysis of Hashimoto who showed for zero CKM4 mixing
that $m_H \, \gtap\, m_{Q_4}$ must be approximately satisfied to
assure the existence of a region between the fourth family quark mass
scale and the scale of new physics in which SM4 is a valid effective
theory amenable to perturbation theory.\cite{hashimoto} Combined with
the PEW data this result excludes vanishing and very small CKM4
mixing. The generalization of Hashimoto's inequality to nonvanishing
CKM4 mixing could provide significant upper and lower bounds on fourth
family mixing angles, depending on the sign and magnitude of the CKM
angle dependent corrections. The lower bound on CKM4 mixing has a
simple explanation: larger values of \mhsp cause the fit to favor
larger mixing, because increased mixing induces an increase in the
oblique parameter $T$ that offsets the decrease due to larger \mh.

The PEW fits depend of course on the inputs, including the choice of
data set and especially \dalfive, the five flavor hadronic
contribution to the running of $\alpha$ to the $Z$ pole, which is the
dominant uncertainty in \almz. While it is reasonable to consider
alternate inputs, the EWWG inputs continue to be a valid, conservative
choice in view of existing systematic uncertainties. A study using
different inputs\cite{el} evidently favored tighter constraints on
CKM4 mixing, although no explicit limits on mixing angles were
presented. To illustrate the sensitivity of the results to the inputs
we explore alternatives to the EWWG defaults: we consider two recent
determinations of \dalfive, an augmented data set with low energy
measurements, and a reduced set which omits the hadronic asymmetry
measurements. In all cases we find that zero CKM4 mixing is excluded
and that CKM4 mixing of order \thcabsp is allowed at 95\% CL, although
when one of the \dalfivesp choices is applied to the data without
hadronic asymmetries only small regions of the parameter space are
allowed.

In the next section we briefly review the SM3 fit and illustrate the
effect of alternative inputs. We then present the EW and Higgs mass
constraints on CKM4 mixing, including the two loop\cite{rb}
nondecoupling contributions $\propto m_{Q_4}^2$ to both $T$ and the
$Z \overline b b$ vertex. For simplicity we assume 3-4 mixing is dominant; the
straightforward generalization to also include 2-4 or 1-4 mixing was
given in our previous work.\cite{mc1} We conclude with a brief
discussion of some of the many aspects of the SM4 scenario that remain
to be explored.

{\it \noindent \underline{SM Fits}}

In this section we compare our SM3 fit to the most recent SM fit of
the EWWG and then compare the impact of various alternative inputs.
In particular we consider two alternates for \dalfivesp and vary the
data set by adding low energy measurements to the ``high energy'' 
set of the EWWG or by removing the hadronic asymmetry measurements as
suggested by one possible interpretation of the \afbbsp anomaly
reviewed below.

\begin{table}
\begin{center}
\vskip 12pt
\begin{tabular}{c|c|cc|cc}
\hline
\hline
 &Experiment& {\bf EWWG} & Pull &{\bf I} & Pull\\
\hline
$\Delta \alpha^{(5)}(m_Z)$ & 0.02758 (35) &0.02768& -0.3 & 0.2768 & -0.3\\
$m_t$ & 173.1 (1.3) &173.2  &-0.1&173.3  &-0.1 \\
$\alpha_S(m_Z)$ &    &0.1185& &0.1180 \\
$m_H$ & & 87 && 89 \\
\hline
$\chi^2$/dof& & 17.3/12 && 17.3/12 \\
CL($\chi^2)$ & & 0.14 &&0.14 \\
\hline
$m_H(95\%)$ & & 155 && 150 \\
CL($m_H\ > 114$ GeV)& & & & 0.23 & \\
\hline
$A_{LR}$ & 0.1513 (21)  & 0.1481  & 1.5& 0.14804 & 1.55 \\
$A_{FB}^l$ & 0.01714 (95) &0.0165  & 0.7&0.1644  &0.7 \\
$A_{e,\tau}$ & 0.1465 (33) & 0.1481 & -0.5& 0.14804 & -0.5 \\
$A_{FB}^b$ & 0.0992 (16) & 0.1038 & -2.9&0.1038&-2.9 \\
$A_{FB}^c$ & 0.0707 (35) & 0.0742 & -1.0&0.0742&-1.0 \\
$Q_{FB}$ & 0.23240 (120) & 0.23138 & 0.8&0.23139&0.8 \\
$\Gamma_Z$ & 2495.2 (23) & 2495.9 &-0.3& 2495.7 & -0.2 \\
$R_{\ell}$ & 20.767 (25) &20.742  & 1.0 &20.739 & 1.1 \\
$\sigma_h$ & 41.540 (37) & 41.478 &1.7  & 41.481 &1.6 \\
$R_b$ & 0.21629 (66) & 0.21579 &0.8& 0.21582 &0.7 \\
$R_c$ & 0.1721 (30) & 0.1723 &-0.1& 0.1722 &-0.04 \\
$A_b$ & 0.923 (20) & 0.935 &-0.6 & 0.935 &-0.6 \\
$A_c$ & 0.670 (27) &  0.668 & 0.1&  0.668& 0.07 \\
$m_W$ & 80.399 (23) & 80.379 & 0.9 & 80.378 & 0.9 \\
\hline
\hline
\end{tabular}
\end{center}
\caption{Comparison of our SM fit with EWWG defaults to the summer 
2009 EWWG fit.\cite{ewwgs09} The 95\% upper limit on \mhsp for the EWWG 
fit reflects the ``blue band'' systematic uncertainties while our value 
does not.}
\end{table}

Table 1 compares our SM fit with EWWG inputs,\footnote{We omit
  $\Gamma_W$; with 2.5\% uncertainty it does not approach the part per
  mil accuracy typical of the other precision data and has little
  effect on the fit.} labeled fit {\bf I}, to the experimental data
and to the EWWG summer 2009 fit.\cite{ewwgs09} The first four rows
display the parameters which are scanned in the fits and are the
inputs to the calculation of the radiative corrections of the other
observables.\footnote{$G_F$ and $m_Z$ are also inputs to the radiative
  corrections. Because they are known much more precisely than the
  other quantities the fits do not change if they are scanned
  and so they are just set to their central values.} We follow the EWWG
practice of allowing the strong coupling constant \alssp to float
unconstrained; the fits are not very sensitive to its precise value as
long as it is within $\sim 2\%$ of 0.118, as for instance in the PDG
fit of \als.\cite{pdg08} The \chisqsp values and confidence levels
are shown in the next two rows. The two fits are virtually identical. 

The acceptable but somewhat marginal 14\% CL in table 1 is due to the
3.2$\sigma$ discrepancy between the two most precise determinations of
\sinthlsp from \afbbsp and \alr, which are seen to have pulls of
opposite sign. The largest pull, 2.9$\sigma$, is borne by \afbbsp
because of an ``alliance'' of \alrsp and the other leptonic asymmetry
measurements with $m_W$, which prefer values of \mhsp near 50 GeV,
while \afbbsp and the other hadronic asymmetry measurements favor $m_H
\simeq 500$ GeV. A possible explanation that cannot
be excluded {\it a priori} is that the hadronic asymmetries 
have underestimated systematic uncertainties,\cite{mcafbb} for which a
leading candidate is the merging, estimated by hadronic Monte Carlo,
of the large QCD radiative corrections with the experimental
acceptance\cite{hfwg} (see the talk cited at \cite{mcafbb}) for a
recent discussion). The anomaly could then be a signal of new physics
to raise the predicted value of the Higgs mass above the 50 GeV scale
predicted in the SM by the leptonic asymmetries and $m_W$.

\begin{table}[t]
\begin{center}
\vskip 12pt
\begin{tabular}{c|c|c|c}
\hline
\hline
 & {\bf I} & {\bf II} & {\bf III} \\
\hline
$\Delta \alpha^{(5)}(m_Z)$ input & 0.02758 (35) &0.02760 (15) & 0.02793(11)\\
\hline
$\Delta \alpha^{(5)}(m_Z)$  & 0.02768 &0.02761 & 0.02797 \\
$m_t$  & 173.3 & 173.3 & 173.3 \\
$\alpha_S(m_Z)$ & 0.118 & 0.118 & 0.1186 \\
$m_H$ & 89 & 94 & 74 \\
\hline
$\chi^2$/dof & 17.3/12 & 17.3/12 & 16.8/12 \\
CL($\chi^2)$ & 0.14 & 0.14 & 0.16 \\
\hline
$m_H(95\%)$ & 150 & 146 & 123 \\
${\rm CL}(m_H \geq 114\ {\rm GeV})$ & 0.23 & 0.22 & 0.08 \\
\hline
\hline
\end{tabular}
\end{center}
\caption{Comparison of SM3 fits to the EWWG data set for 
three input values of $\Delta \alpha^{(5)}(m_Z)$. }
\end{table}

\begin{figure}[t]
\centerline{\includegraphics[width=4in,angle=90]{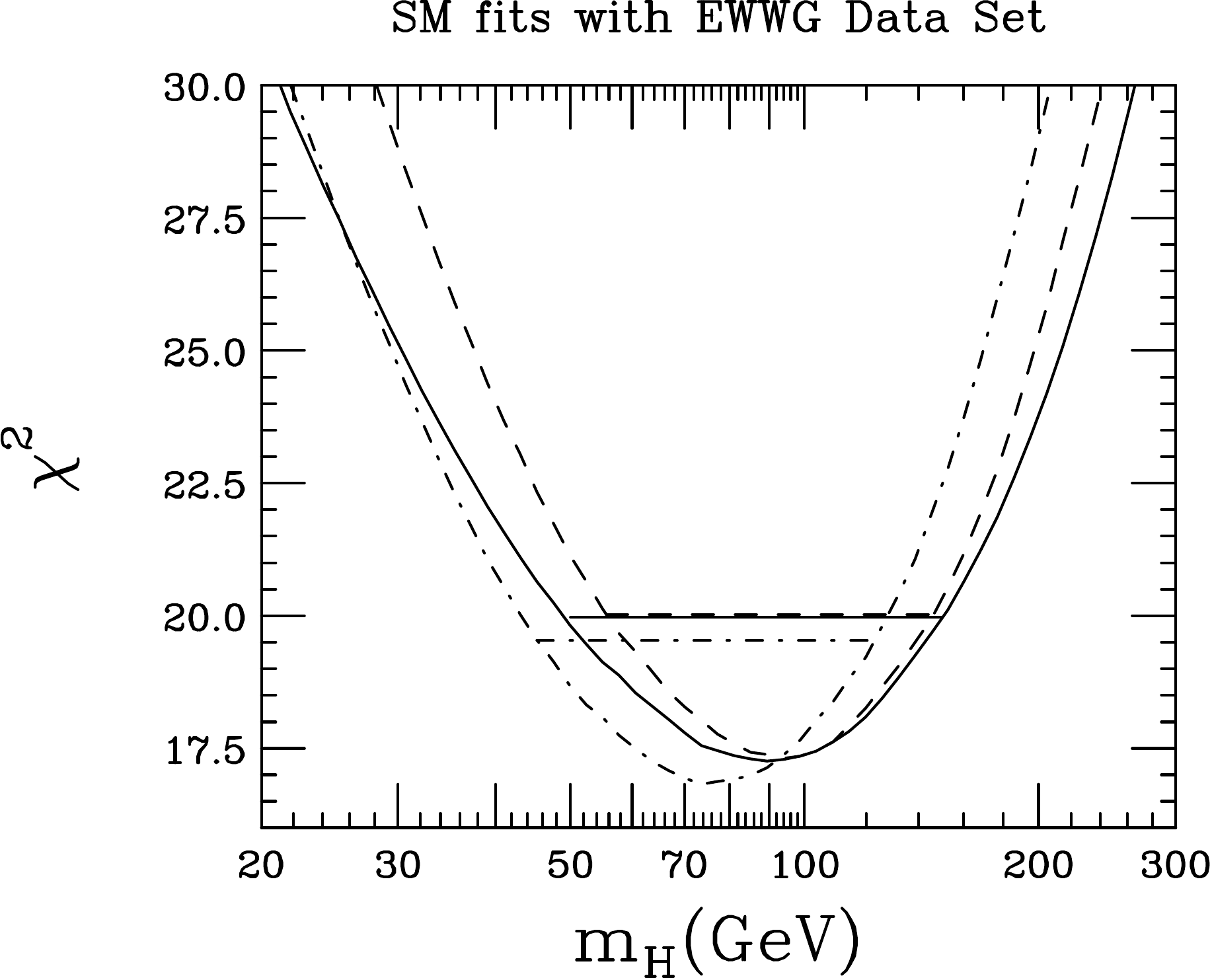}}
\caption{$\chi^2(m_H)$ for the SM fits in table 2: {\bf I} solid, {\bf
    II} dashed, and {\bf III} dash-dot.  The horizontal lines indicate
  the symmetric 90\% confidence intervals. }
\label{fig1}
\end{figure}

To explore the effect of alternative inputs we begin by considering
the EWWG data set of table 1 but with two different, recently obtained
values for \dalfive. The results are summarized in table 2, where we
display the input values for \dalfivesp and the corresponding best fit
results for the four scanned parameters (\dalfive, $m_t$, \als, and
\mh) that control the radiative corrections. Fit {\bf I} is the same
fit shown in table 1 with the EWWG default \dalfive.\cite{bp} The fit
{\bf II} value of \dalfive\cite{hagiwara} uses a BABAR analysis of
radiative return data to determine $\sigma(e^+e^- \to {\rm hadrons})$
in the poorly known region between 1.4 and 2 GeV. The result is almost
identical to the EWWG default but with a much smaller quoted
error. The fit {\bf III} value of \dalfive,\cite{erler} which was used
in the SM4 fits reported in \cite{el}, uses tau decay data as well as
the BABAR data. The resulting \dalfivesp has a larger central value
and a much smaller quoted error than the EWWG default.\footnote{In
  \cite{erler} the three-flavor contribution $\Delta \alpha^{(3)}$ is
  computed and then RG extrapolated to the $Z$-pole.  To a good
  approximation the result corresponds to the value of \dalfivesp
  given in table 2.\cite{erler-private}} The values of \dalfivesp in
{\bf II} and {\bf III} differ by almost $2\sigma$. Fits {\bf I} and
{\bf II} in table 2 are nearly identical, indicating that the
increased precision of the fit {\bf II} \dalfivesp does not have a
major impact on the best fit.  The higher central value of \dalfivesp
in fit {\bf III} does have an appreciable effect, pulling \mhsp to
smaller values, with the 95\% CL upper limit at 123 GeV, and creating
tension with the LEPII 114 GeV lower limit, with which it is
inconsistent at 92\% CL.  Figure 1 shows \chisqsp plotted as a
function of \mhsp for the three fits, with 90\% symmetric confidence
intervals indicated by the horizontal lines.

Next we consider the effect of adding low energy data to the EWWG data
set. There are three candidates that might appreciably affect the fit
of \mh: the weak charge $Q_W$ of the $^{133}{\rm Cs}$ nucleus measured
in atomic parity violation and the measurements of \sinthwsp in
M\"oller scattering at SLAC\cite{moller} and in $\nu N$ scattering by
the NuTeV collaboration.\cite{nutev} In the past the EWWG has included
the APV and NuTeV measurements but eventually omitted them from the
fit because of concerns about systematics. Today the original NuTeV
result is no longer relevant because of a subsequent measurement by
the NuTeV collaboration of an asymmetry in the nucleon $\overline ss$
sea\cite{ntssbar} as well as changes in several other related
measurements. For the fits in \cite{el} the NuTeV result was revised
with four modifications of the original analysis; we prefer to omit
the NuTeV measurement pending an authoritative analysis by the NuTeV
collaboration itself. In any case the omission has little effect on
the results.

A recent calculation\cite{APV} of Cesium atomic transistions claims a
significantly reduced theoretical uncertainty and the central value,
which has previously disagreed with the SM by $2\sigma$, now agrees
almost precisely (with $Q_W = -73.16 \pm 0.29 \pm 0.20$ while our SM
fit yields $Q_W = -73.14$) and therefore contributes zero to the
\chisq. The error estimate is tested by comparing calculations of
several supporting quantities with experiment.\cite{APV2} We
include this new APV result and also the M\"oller scattering
measurement, which translated to the $Z$-pole is ${\rm sin}^2
\theta_W^{\ell\ {\rm eff}}= 0.23339 \pm 0.00140$.\cite{pdg08} It
provides a less precise value of \sinthlsp than \qfb, which is the
least precise of the six $Z$ decay asymmetry measurements in table
1. In the SM fits the M\"oller measurement has a pull of $1.4\sigma$
and tends to slightly increase \mh.

\begin{table}[t]
\begin{center}
\vskip 12pt
\begin{tabular}{c|c|c|c}
\hline
\hline
 & {\bf I} & {\bf II} & {\bf III} \\
\hline
$\Delta \alpha^{(5)}(m_Z)$ input & 0.02758 (35) &0.02760 (15) & 0.02793(11)\\
\hline
$\Delta \alpha^{(5)}(m_Z)$  & 0.02768 &0.02761 & 0.02797 \\
$m_t$  & 173.3 & 173.3 & 173.3 \\
$\alpha_S(m_Z)$ & 0.118 & 0.118 & 0.1186 \\
$m_H$ & 94 & 99 & 74 \\
\hline
$\chi^2$/dof & 19.3/14 & 19.3/14 & 18.8/14 \\
CL($\chi^2)$ & 0.16 & 0.16 & 0.17 \\
\hline
$m_H(95\%)$ & 154 & 150 & 127 \\
${\rm CL}(m_H \geq 114 \ {\rm GeV})$ & 0.25 & 0.25 & 0.10 \\
\hline
\hline
\end{tabular}
\end{center}
\caption{Comparison of SM3 fits to the EWWG data set plus low 
energy measurements ($Q_W(Cs)$ and M\"oller scattering) for three input 
values of $\Delta \alpha^{(5)}(m_Z)$. }
\end{table}

Fits including the low energy measurements are shown in table 3.  The
low energy data has little effect except for the very small increase
in \mh. The \chisqsp increases by two for the two additional degrees
of freedom and the \chisqsp confidence levels hardly change.  Fit {\bf
  III} continues to exhibit tension with the LEPII lower limit on
\mh. If we include the NuTeV measurement as revised in \cite{el} the
only effect on the fits is another very slight increase in \mh, e.g.,
for fit III the 95\% CL upper limit increases from 127 to 130 GeV and
the CL for $m_H\ >\ 114$ GeV increases from 0.10 to 0.11.

\begin{table}[t]
\begin{center}
\vskip 12pt
\begin{tabular}{c|c|c|c}
\hline
\hline
 & {\bf I} & {\bf II} & {\bf III} \\
\hline
$\Delta \alpha^{(5)}(m_Z)$ input & 0.02758 (35) &0.02760 (15) & 0.02793(11)\\
\hline
$\Delta \alpha^{(5)}(m_Z)$  & 0.02761 &0.02761 & 0.02790 \\
$m_t$  & 173.3 & 173.3 & 173.3 \\
$\alpha_S(m_Z)$ & 0.118 & 0.118 & 0.118 \\
$m_H$ & 52 & 52 & 45 \\
\hline
$\chi^2$/dof & 5.59/9 & 5.59/9 & 5.76/9 \\
CL($\chi^2)$ & 0.78 & 0.78 & 0.76 \\
\hline
$m_H(95\%)$ & 105 & 94 & 80 \\
${\rm CL}(m_H \geq 114\ {\rm GeV})$ & 0.030 & 0.016 & 0.0035 \\
\hline
\hline
\end{tabular}
\end{center}
\caption{Comparison of SM3 fits to the EWWG data set minus the 
three front-back hadronic asymmetry measurements for three input 
values of $\Delta \alpha^{(5)}(m_Z)$. }
\end{table}

Table 4 shows the effect of removing the three hadronic asymmetry
measurements from the EWWG data set, as would be appropriate if the
\afbbsp anomaly results from underestimated systematic
error.\cite{mcafbb} The \chisqsp CL's increase to robust levels but
the \mhsp predictions fall significantly below the LEPII lower
limit. The conflict is most acute for fit {\bf III}, which is
inconsistent with the LEPII limit at 99.65\% CL. Conflict between the fit
and the direct limit could be a signal of new physics to
increase the \mhsp prediction, as occurs generically for new physics
models with $T\ >\ 0$, for example, in SM4.

To summarize this section, the greatest changes to the SM fit from the
alternatives to the EWWG defaults that we have considered arise from
the \dalfivesp of fit {\bf III} and from the exclusion of the hadronic
asymmetry measurements, while the increased precision of the fit {\bf
  II} \dalfivesp and the addition of the low energy measurements have
less impact.

{\it \noindent \underline{SM4 Fits and Higgs mass constraints}}

We now consider the correlated constraints on CKM4 mixing that result
from the combination of recent SM4 Higgs mass constraints with the
constraints from the precision EW data. We assume for simplicity that
the fourth family mixes predominantly with the third and will exhibit
the 95\% CL allowed regions in the $s_{34}-m_H$ plane, where $s_{34} =
{\rm sin}\theta_{34}$ is the sine of both the $t^{\prime}-b$ and
$b^{\prime}-t$ mixing angles. Following \cite{el} we choose
$m_{t^{\prime}} - m_{b^{\prime}} = 16$ GeV and $m_{\tau^{\prime}} -
m_{\nu^{\prime}} = 91$ GeV, yielding a slightly lower \chisqsp (by
$\sim 0.5$) than the masses used in \cite{mc1} which were based on the
fits of \cite{kribs}.  We fix $m_{\nu^{\prime}} = 101$ GeV and
consider two choices for the quark masses, $m_{b^{\prime}} = 338$ GeV,
which is at the current CDF\cite{cdfbprime} lower limit,\footnote{The
  CDF limit on $m_{b^{\prime}}$ assumes ${b^{\prime}} \rightarrow tW$
  is dominant and prompt. For other scenarios other limits will apply
  --- see \cite{otherlimit}.} and $m_{b^{\prime}} = 484$ GeV
corresponding to $m_{t^{\prime}} = 500$ GeV, at the perturbative
unitarity limit.\cite{cfh} In addition to the leading one loop
nondecoupling contributions to $T$ and the $Z \overline bb$ vertex
proportional to $m_{Q_4}^2$ we also include the leading nondecoupling
two loop contributions proportional to $m_{Q_4}^4$.\cite{rb} As shown
in \cite{mc1}, perturbation theory for the nondecoupling corrections
remains under control for $m_{t^{\prime}} = 500$ GeV but has
decisively broken down at $m_{t^{\prime}} = 1$ TeV.  In addition to
the EWWG defaults we consider alternative inputs as in the previous
section.

\begin{table}[t]
\begin{center}
\vskip 12pt
\begin{tabular}{c|c|c|c}
\hline
\hline
 & {\bf I} & {\bf II} & {\bf III} \\
\hline
$\Delta \alpha^{(5)}(m_Z)$ input & 0.02758 (35) &0.02760 (15) & 0.02793(11)\\
\hline
$\Delta \alpha^{(5)}(m_Z)$  & 0.02754 &0.02761 & 0.02790 \\
$m_t$  & 173.3 & 173.3 & 173.3 \\
$\alpha_S(m_Z)$ & 0.1174 & 0.1174 & 0.1174 \\
$m_H$ & 114 & 109 & 89 \\
\hline
$\chi^2$/dof & 16.7/12 & 16.7/12 & 17.0/12 \\
CL($\chi^2)$ & 0.16 & 0.16 & 0.15 \\
\hline
\hline
\end{tabular}
\end{center}
\caption{SM4 fits to the EWWG data set 
with $m_{t^{\prime}},m_{b^{\prime}} = 354,338$ GeV, 
$S,T = 0.139,0.159$ and $s_{34}=0$,
for three values of $\Delta \alpha^{(5)}$.} 

\end{table}

The 95\% CL $s_{34} - m_H$ contours ($\Delta \chi^2 = 5.992$) are
defined with respect to the best fit for each set of fourth family
masses and fit inputs, which always occurs at $s_{34}=0$. The best
fits for the parameter set with $m_{t^{\prime}},m_{b^{\prime}} =
354,338$ GeV are shown in table 5 for the EWWG data set with the three
values of \dalfivesp considered above. The \chisqsp confidence level
is defined in a Bayesian sense, assuming that the fourth family masses
are known {\it a priori}. Comparing with the corresponding SM3 fits in
table 2, we see that the \chisqsp is slightly lower, by $\Delta \chi^2
= -0.6$, for SM4 fits {\bf I} and {\bf II} than for the corresponding
SM3 fits and slightly higher (+0.2) for fit {\bf III}. The difference
probably reflects the strong preference of the fit {\bf III}
\dalfivesp for small \mhsp seen in the SM3 fits of the previous
section, since positive $T$ in the SM4 fits puts ``upward pressure''
on \mh.  This effect is more pronounced in the 95\% contour plots
shown below. The best fits including the low energy measurements are
very similar and are not displayed.  The fits with
$m_{t^{\prime}},m_{b^{\prime}} = 500,484$ GeV and $s_{34}=0$ are also
virtually identical to those of table 5 and are also not displayed.
  
The sensitivity of the fits to $\theta_{34}$ arises from nondecoupling
heavy fermion contributions to the $Z \overline bb$ vertex
correction\cite{yanir,alwall} and, predominantly, to the oblique
parameter $T$, given at one loop by\cite{mc1}
\begin{eqnarray}
T_4= \frac{1}{8\pi x_W(1-x_W)}\left\{ 3\left[
     F_{t^{\prime}b^{\prime}} +s_{34}^2(F_{t^{\prime}b} 
     +F_{tb^{\prime}} -F_{tb} -F_{t^{\prime}b^{\prime}})\right] 
     + F_{l_4\nu_4}\right\}.
\end{eqnarray}
and
\begin{eqnarray}
\delta^V g_{bL}^{3-4} = s_{34}^2\, \frac{\alpha}{16\pi x_W(1-x_W)}\, 
     \left(\frac{m_{t^{\prime}}^2}{m_Z^2} - \frac{m_t^2}{m_Z^2}\right)
\end{eqnarray}
where $x_i = m_i^2/m_Z^2$, $x_W={\rm sin}\theta_W^2$, and 
\begin{eqnarray}
   F_{12}= \frac{x_1 +x_2}{2}
              -\frac{x_1x_2}{x_1-x_2}{\rm ln}\frac{x_1}{x_2}
\end{eqnarray}
is the well known nondecoupling fermionic
correction to the rho parameter.\cite{veltman,cfh}

\begin{figure}[t]
\centerline{\includegraphics[width=3.75in,angle=90]{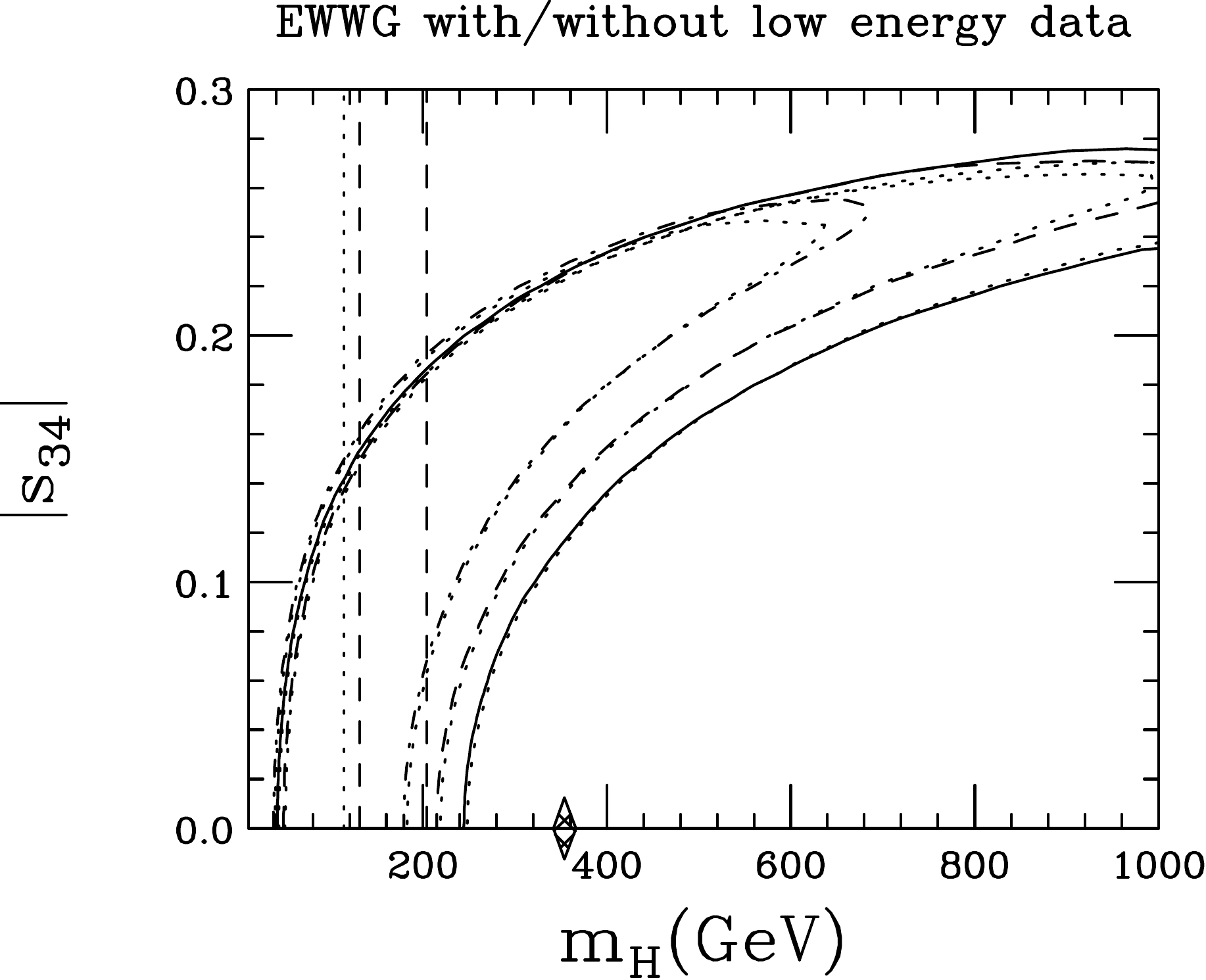}}
\caption{95\% CL contour plots for SM4 with
  $m_{t^{\prime}},m_{b^{\prime}} = 354,338$ GeV: {\bf I} solid, {\bf
    II} dashed, and {\bf III} dash-dot. The nearly identical dotted
   contours include the low energy data. The vertical dotted and dashed
  lines indicate the LEPII and Tevatron 95\% exclusion regions, and
  the diamond on the abscissa marks the stability bound
   $m_H \gtrsim m_{t^{\prime}}$.}
\label{fig2}
\end{figure}

Figures 2 and 3 display the 95\% CL contours for $m_{t^{\prime}} =
354$ and $m_{t^{\prime}} = 500$ GeV respectively, for the EWWG data
set both with and without the addition of the low energy measurements,
and for the three values of \dalfivesp considered previously. The
contour bounded by the solid line corresponds to the EWWG default data
set and default value for \dalfive, labeled as fit {\bf I} in table
5. The dotted line that lies just within the solid contour is the
result of adding the low energy measurements to the data set, seen to
have an almost negligible effect.  Similarly the dashed and dot-dashed
contours and their accompanying dotted lines are the result of the fit
{\bf II} and fit {\bf III} \dalfivesp values respectively. As \mhsp
increases above 250 GeV the fits imply both lower and upper limits on
$|s_{34}|$.  For instance, at $m_H=m_{t^{\prime}} = 354$ GeV, the fit
{\bf I} contour restricts $\theta_{34}$ to the interval $0.115 \leq
|s_{34}| \leq 0.225$. Similarly for fit {\bf I} at $m_H=m_{t^{\prime}}
= 500$ GeV, $\theta_{34}$ is restricted to $0.07 \leq |s_{34}| \leq
0.14$.  As was the case for the SM3 fits, we see that the fit {\bf I}
and fit {\bf II} contours are very similar, with nearly identical
upper limits on $|s_{34}|$ although fit {\bf II} yields a more
restrictive lower limit. In both cases the reach in \mhsp extends to 1
TeV and the upper limit on $s_{34}$ extends to $\simeq 0.26$ for
$m_{t^{\prime}} = 354$ GeV and to $\simeq 0.17$ for $m_{t^{\prime}} =
500$ GeV.

\begin{figure}[t]
\centerline{\includegraphics[width=3.75in,angle=90]{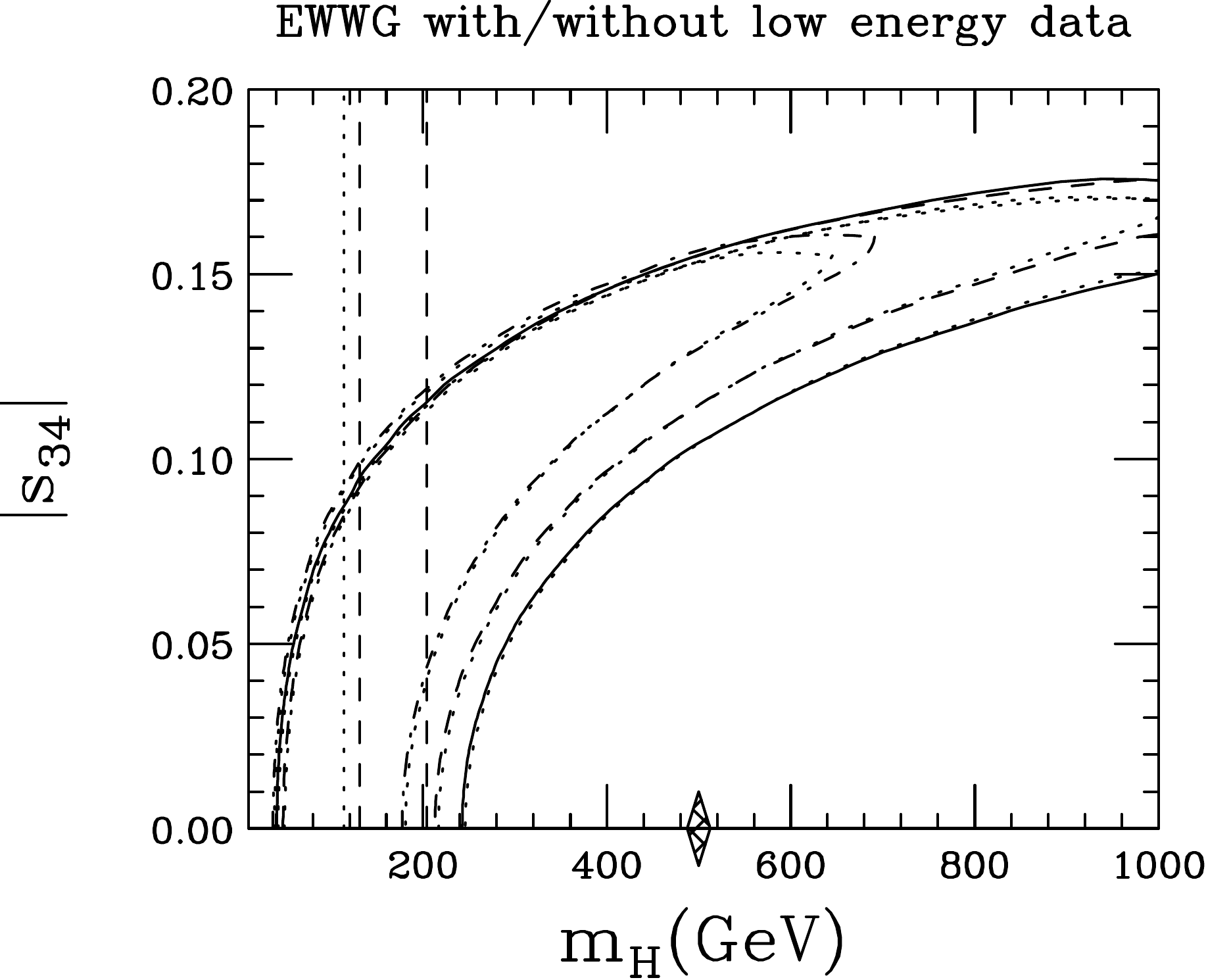}}
\caption{95\% CL contour plots for SM4 with
  $m_{t^{\prime}},m_{b^{\prime}} = 500,484$ GeV, as in figure 2.}
\label{fig3}
\end{figure}

As was the case for the SM3 fits the most significant effect of the
alternative inputs is from the fit {\bf III} value of \dalfive. The
fit {\bf III} contours for the upper limit of $|s_{34}|$ follow those
of fits {\bf I} and {\bf II} quite closely until $m_H=690$ (660) GeV,
for figure 1 (2), which shows that the somewhat tighter upper limit on
$|s_{34}|$ in fit {\bf III} is a consequence of the well known
correlation between larger \dalfivesp and smaller values of \mhsp that
we saw in the SM3 fits of the previous section. For $m_H\ \ltap\ 600$
GeV the upper limit on $|s_{34}|$ is nearly the same for all three
fits although the lower limits from fits {\bf II} and {\bf III} are
stronger. In particular, for $m_H=m_{t^{\prime}} = 354$ GeV and
$m_H=m_{t^{\prime}} = 500$ GeV, the upper limit on $|s_{34}|$ is
nearly the same for all three \dalfivesp inputs, with or without 
the low energy data.

The direct experimental limits on \mhsp are indicated by the vertical
lines in figures 2 and 3, and a theoretical limit is indicated by the
diamond on the abscissa at $m_H=m_{t^{\prime}}$. The dotted vertical
line denotes the LEPII 95\% CL lower limit and the region between the
two dashed vertical lines marks the 131 -- 204 GeV SM4 95\% CL
exclusion region established by CDF and D0. Except for the interval
between 114 and 131 GeV we see that $\theta_{34} = 0$ is excluded by
fit {\bf III} and is nearly excluded by fit {\bf II}, and that fit
{\bf III} requires $0.255 > |s_{34}| > 0.065$ for $m_{t^{\prime}} =
354$ GeV and $0.16 > |s_{34}| > 0.035$ for $m_{t^{\prime}} = 500$
GeV. The lower limits on $|s_{34}|$ will be strengthened as the
limits on the SM4 Higgs boson are tightened at the Tevatron and the
LHC (or when it is discovered!).

The theoretical lower limit on \mhsp indicated by the diamond in
figures 2 and 3 at $m_H=m_{t^{\prime}}$ applies strictly speaking only
at $\theta_{34} = 0$, but it is safe to say that it also excludes at
least some small region around $\theta_{34} = 0$. The bound
$m_H\ \gtap\ m_{t^{\prime}}$ follows from the assumption that the
cutoff for new BSM4 physics is no lower than 2 TeV, and can be relaxed
to $m_H\ \gtap\ m_{t^{\prime}} - 50$ GeV if the cutoff is lowered to 1
TeV.\cite{hashimoto} Since the analysis of \cite{hashimoto} was
performed assuming $\theta_{34} = 0$, we cannot be sure how it affects
the allowed region of the entire $|s_{34}| - m_H$ plane. Assuming that
the bound for $\theta_{34} \neq 0$ is of the form
$m_H\ \gtap\ m_{t^{\prime}}(1 + C|s_{34}|)$, the sign and value of the
coefficient $C$ will determine the further restrictions on $|s_{34}|$;
if it is of the form $m_H\ \gtap\ m_{t^{\prime}}(1 + C|s_{34}|^2)$ the
corrections will be small.  If the $s_{34}$ correction is quadratic or
if $C$ is very small the entire region $m_H\ \ltap\ m_{t^{\prime}}$
would be excluded and the allowed region would establish lower and
upper limits on $|s_{34}|$, which for the EWWG defaults, fit {\bf I},
would be $0.275 > |s_{34}| > 0.115$ for $m_{t^{\prime}} = 354$ GeV and
$0.175 > |s_{34}| > 0.105$ for $m_{t^{\prime}} = 500$ GeV.

\begin{figure}
\centerline{\includegraphics[width=3.75in,angle=90]{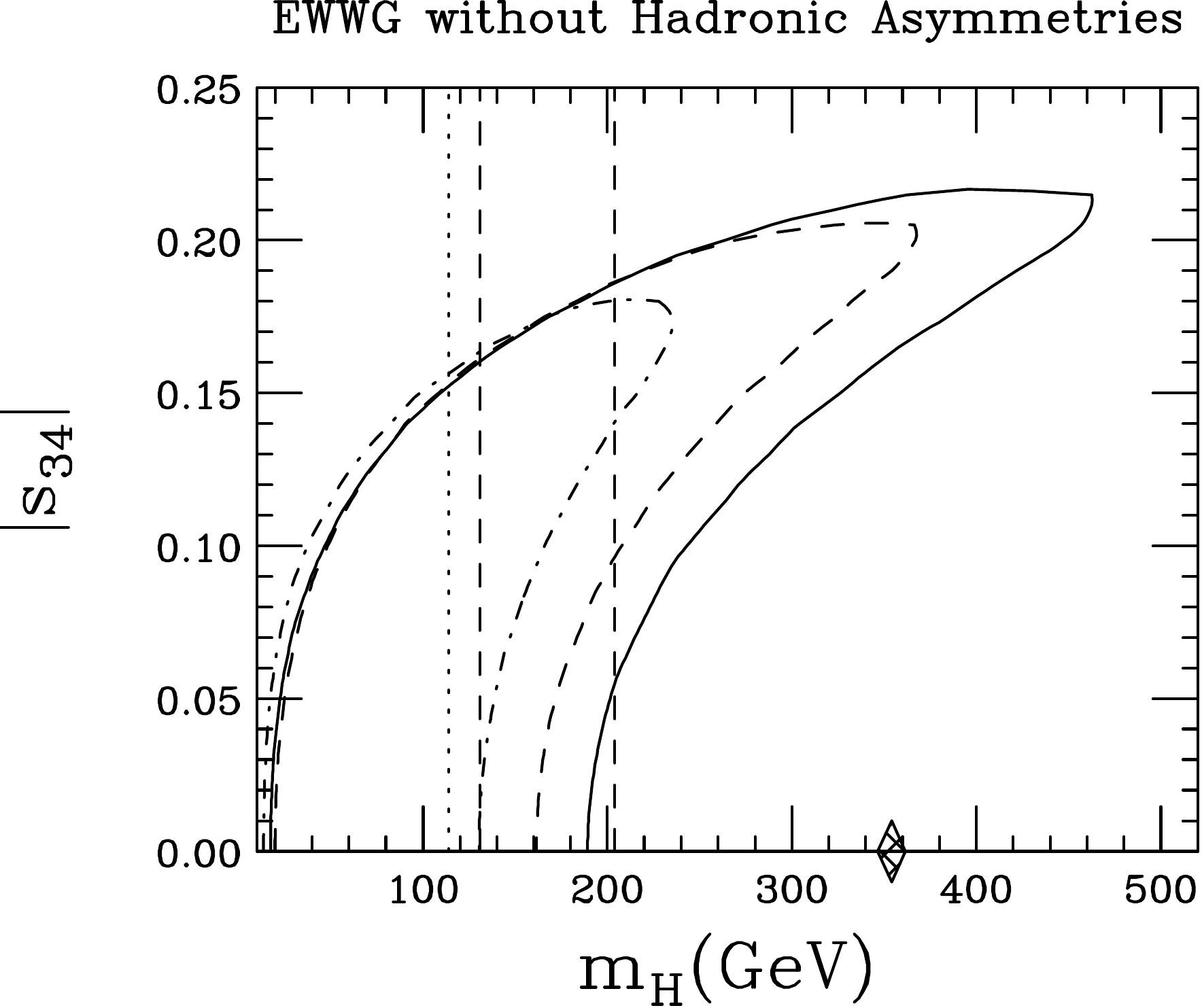}}
\caption{SM4 95\% CL contour plots with $m_{t^{\prime}},m_{b^{\prime}}
  = 354,338$ GeV, as in figure 2, without the three hadronic asymmetry
  measurements.}
\label{fig4}
\end{figure}
\begin{figure}
\centerline{\includegraphics[width=3.75in,angle=90]{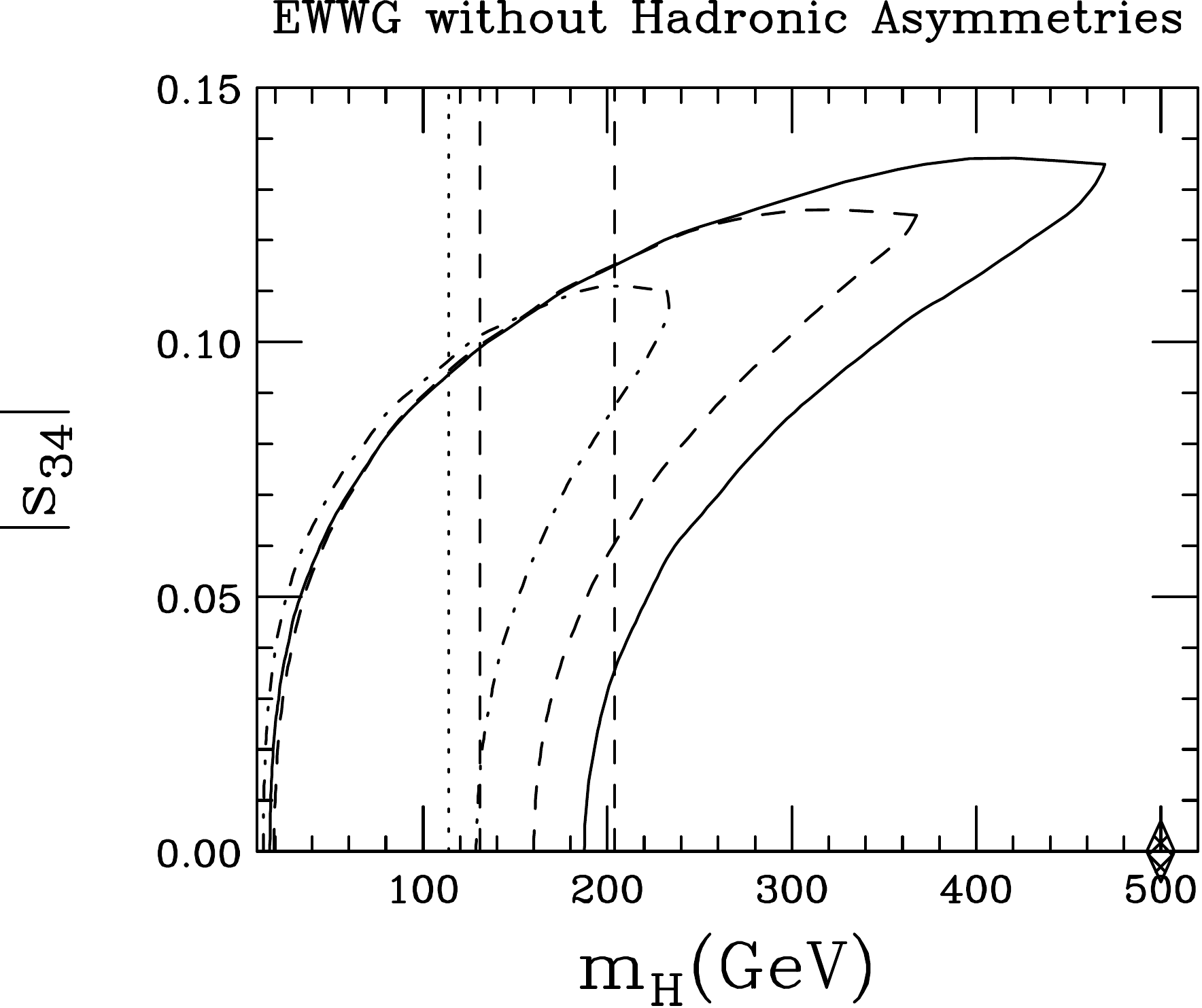}}
\caption{SM4 95\% CL contour plots with $m_{t^{\prime}},m_{b^{\prime}}
  = 500,484$ GeV, as in figure 2, without the three hadronic asymmetry
  measurements.}
\label{fig5}
\end{figure}

Figures 4 and 5 show the contour plots for the data set consisting of
the EWWG ``high energy'' measurements listed in table 1 but without
the three hadronic asymmetry measurements, \afbb, \afbc, and \qfb, as
would be appropriate if the \afbbsp anomaly is due to underestimated
systematic error. The corresponding SM3 fits were summarized in table
4. We see from figures 4 and 5 that SM4 removes the tension with the
114 GeV LEPII lower limit on \mhsp that is apparent in table 4, but
that in the case of fit {\bf III} the CDF-D0 limit on the SM4 Higgs
boson mass excludes most of the remaining allowed parameter space
except for two small regions between 114 and 131 GeV and between 204
and 230 GeV. Fits {\bf I} and {\bf II} have extensive allowed regions
above $m_H=204$ GeV but depending on how mixing affects the
theoretical lower limit on \mh, i.e., the parameter $C$ discussed
above, they could be severely constrained or even entirely excluded by
the theoretical limit.

\begin{figure}
\centerline{\includegraphics[width=3.75in,angle=90]{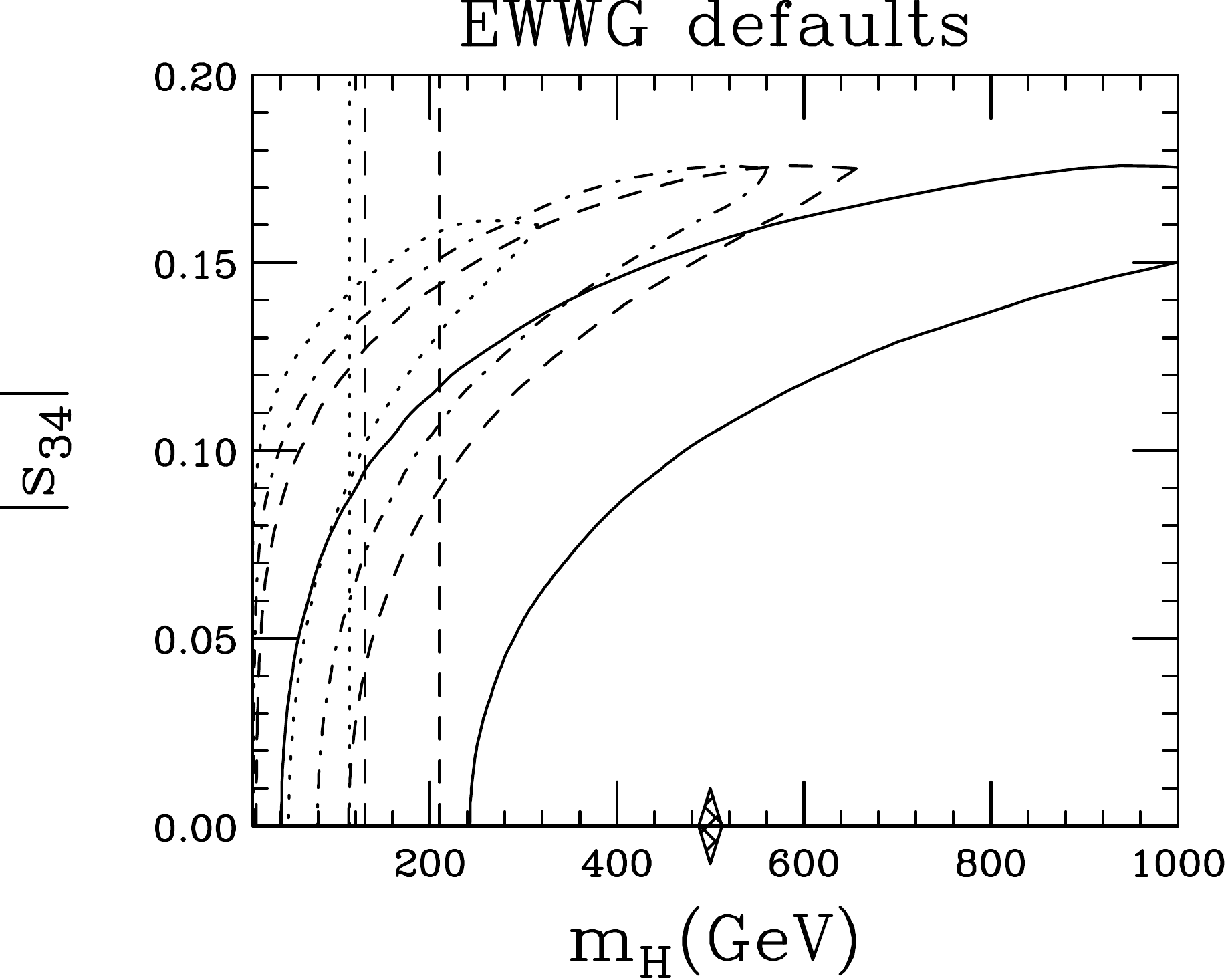}}
\caption{95\% CL contours as in figure 2 using EWWG defaults, all with
  $m_{t^{\prime}} = 500$ GeV and $m_{\nu^{\prime}} = 100$ GeV. For the
  dotted and dash-dotted contours $m_{b^{\prime}} = 500$ GeV while
  $m_{\ell^{\prime}} - m_{\nu^{\prime}} = 0$ and 40 GeV
  respectively. The dashed contour is for the best fit model described
  in the text, with $m_{b^{\prime}} = 475$ GeV and $m_{\ell^{\prime}}
  - m_{\nu^{\prime}} = 45$ GeV. The solid contour, reproduced from
  figure 3, is for $m_{b^{\prime}} = 484$ GeV and $m_{\ell^{\prime}} -
  m_{\nu^{\prime}} = 91$ GeV.}
\label{fig6}
\end{figure}

The quark and lepton masses considered above were chosen because they
yield good fits for $s_{34}=0$. We may then ask whether different
masses might provide better fits for $s_{34}\neq 0$. 
Since the decrease in confidence level for increasing $|s_{34}|$ is a
consequence of the associated increase in $T$, it might seem that
quark and lepton doublets with smaller mass splittings could yield
improved fits at large $|s_{34}|$ even if they are not favored at
$s_{34}=0$.  We find however that this strategy does not increase the
upper limit on $|s_{34}|$ although it does change the relationship
between $s_{34}$ and \mh. The quark mass splitting used above,
$m_{t^{\prime}} - m_{b^{\prime}} = 16$ GeV, results in a very small
value of $F_{t^{\prime}b^{\prime}}$, two orders of magnitude smaller
than $F_{tb}$ and three orders of magnitude smaller than
$F_{t^{\prime}b}$ for $m_{t^{\prime}}=500$ GeV, so that the resulting
fits are nearly identical to those with $m_{t^{\prime}} =
m_{b^{\prime}}$.  Thus the only possibility to pursue this strategy is
to reduce the lepton mass splitting from the 91 GeV value used in the
fits presented above.

We illustrate this strategy with two models, taking $m_{t^{\prime}} =
m_{b^{\prime}} = 500$ GeV, $m_{\nu^{\prime}} = 100$ GeV and
$m_{\ell^{\prime}} - m_{\nu^{\prime}} = 0$ or 40 GeV. Unlike the cases
considered above, in these models \chisqsp does not increase
monotonically from a minimum at $s_{34}=0$.  Instead the \chisqsp
minimum occurs at $|s_{34}| \simeq 0.06$ but at unacceptably small
Higgs masses, around 30 to 40 GeV. The 95\% CL contours are shown in
figure 6 (dots and dash-dots) where they are compared to the model
with $m_{\ell^{\prime}} - m_{\nu^{\prime}} = 91$ GeV (solid). We see
that the upper limit on $|s_{34}|$ is not increased and that the
allowed parameter space is significantly reduced by the Higgs mass
constraints. The preference for small \mhsp in these fits is a
consequence of the interplay between the oblique parameters $S$ and
$T$ with the Higgs mass: decreasing the mass splittings reduces $T$
and increases $S$, which both drive the fit to small \mh. Though the
parameter space is reduced these models are currently still
viable. They do not achieve larger upper limits on $|s_{34}|$, but 
they do allow larger mixing at smaller \mh.

Finally we consider the masses that yield the lowest \chisqsp minimum
we have obtained using EWWG defaults.  Fixing $m_{t^{\prime}}=500$ GeV
and $m_{\nu^{\prime}}=100$ GeV, we find that the smallest \chisqsp
occurs for $m_{t^{\prime}}-m_{b^{\prime}} \simeq 25$ GeV,
$m_{\ell^{\prime}}-m_{\nu^{\prime}}\simeq 40$ GeV, and $s_{34}=0$.
The value at the minimum, $\chi^2=15.9$, is 1.4 units less than the
SM3 best fit, for which the minimum is $\chi^2=17.3$ (see table
1). While this model has the deepest \chisqsp minimum we have found
and that minimum occurs at $s_{34}=0$, the upper limit $|s_{34}|\,
\ltap \, 0.175$ is as large as that of any of the other models, as seen 
in the dashed contour in figure 6.  The allowed range of CKM4 mixing
is not extended by tuning masses to move the \chisqsp minimum to
$s_{34}\neq 0$.

{\it \noindent \underline{Discussion}}

We have explored the constraints on fourth family CKM mixing that
arise from the combined effect of the precision electroweak data and
bounds on the mass of the SM4 Higgs boson. We find that fourth family
CKM mixing of order \thcabsp is allowed, with $|s_{34}|\ \ltap\ 0.27$
for quark masses near the Tevatron lower limit and with
$|s_{34}|\ \ltap\ 0.17$ for masses near the perturbative unitarity
upper limit, and that the experimental and theoretical bounds on \mhsp
in SM4 favor nonvanishing fourth family CKM mixing.  In addition to
the EWWG default inputs we explored alternate data sets and values of
\dalfive.  Adding low energy data has little effect but removing the
hadronic front-back asymmetries leads to much tighter constraints.
One recent determination\cite{hagiwara} of \dalfivesp yields very
similar results to the EWWG default fits while another\cite{erler,el}
strongly prefers small \mh, causing tension with the LEPII 114 GeV
lower limit in SM3 and a tighter upper limit on \mhsp in SM4. A
possible explanation of the difference is that \cite{erler} uses 
$\tau$ decay data, while \cite{hagiwara} does not because of concern
over systematic uncertainties associated with the isospin corrections
required in the $\tau$ analysis. It would be interesting to know the
extent to which the $\tau$ data influences the result in \cite{erler}.

This work is restricted to the implications of the precision EW data
and the Higgs mass limits on CKM mixing in SM4. In a truly global
approach data from two other areas should also be confronted. Lacker
and Menzel have made the interesting observation that the extraction
of $G_F$ in SM4 leaves room for significant deviations from its value
in SM3 which would affect the analysis of the precision EW data, and
that the CKM4 and PMNS4 matrices should be considered simultaneously
in a single fit.\cite{lm} The EW fit of the CKM4 matrix can also be
extended by including FCNC and CP constraints together with the EW
constraints, and a first step in this direction has been
taken.\cite{lenz2} Both are clearly important directions to pursue.

The effect of the Higgs boson mass bounds on the allowed CKM4
parameter space is already significant, as shown in the results
reported above. They will become more powerful as the Higgs boson
searches at the Tevatron and LHC progress and when the RG analysis of
the stability of the SM4 Higgs potential is generalized to account for
fourth family CKM mixing. Eventually they will tighten the mixing
constraints or even exclude the {\em perturbative} SM4 scenario.
Although \thcabsp order mixing is allowed within the 95\% CL contours,
the minimum \chisqsp is at $\theta_{34}=0$ and, as seen in
\cite{mc1}, the confidence levels of the best EW fits decrease
monotonically as $|s_{34}|$ increases.

Of course the underlying assumption of both the EW fits and the RG
analysis of the Higgs potential is that SM4 exists as an effective
field theory that can be approximately described by perturbation
theory within some energy domain. This is a plausible assumption for
fourth family masses below the perturbative unitarity bounds but fails
for larger masses; in particular, in \cite{mc1} we traced how the two
loop nondecoupling corrections grow as $m_{t^{\prime}}$ increases
above the unitarity bound toward 1 TeV. If there is little or no
hierarchy between the heavy quark threshold at $2 m_{t^{\prime}}$ and
the scale of new strong dynamics, then the EW fits and the RG analysis
both become quantitatively unreliable. In that case  the LHC
will encounter a new realm of strong dynamics whose exploration will
make for a very rich physics program. If the fourth family quarks
are very heavy, e.g., $m_Q\, \gtap\, 1$ TeV, and difficult or
impossible to observe directly, they will give rise to a large $gg
\to ZZ$ (and $WW$) signal in the diboson energy region $m_H^2\, <
s_{ZZ}\, <\, 4m_{Q_4}^2$ that could be seen at the LHC with $5\sigma$
significance over backgrounds with only O(10) fb$^{-1}$ of integrated
luminosity,\cite{mczz} before direct detection would be possible.

\vskip .2in
\noindent {\it Acknowledgements:} I would like to thank Jens Erler for
providing detailed information concerning the fits in \cite{el} and 
Zoltan Ligeti for helpful comments.

\vskip 0.1in
\noindent{\small This work was supported in part by the Director, Office of
Science, Office of High Energy and Nuclear Physics, Division of High
Energy Physics, of the U.S. Department of Energy under Contract
DE-AC02-05CH11231}

\end{document}